\documentclass{article}

     \usepackage[preprint]{neurips_2019}

\usepackage[utf8]{inputenc} 
\usepackage[T1]{fontenc}    
\usepackage{hyperref}       
\usepackage{url}            
\usepackage{booktabs}       
\usepackage{amsfonts}       
\usepackage{nicefrac}       
\usepackage{microtype}      
\usepackage{amsmath}       

\title{Application and Computation of Probabilistic Neural Plasticity}

\author{%
  Soaad Q. Hossain \\
  Department of Computer and Mathematical Sciences, Department of Philosophy\\
  University of Toronto, Scarborough \\
  Toronto, Canada \\
  \texttt{soaad.hossain@mail.utoronto.ca} \\
}

\begin{document}

\maketitle

\begin{abstract}

  The discovery of neural plasticity has proved that throughout the life of a human 
   being, the brain reorganizes itself through forming new neural connections. The 
   formation of new neural connections are achieved through the brain's effort to 
   adapt to new environments or to changes in the existing environment. Despite
   the realization of neural plasticity, there is a lack of understanding the probability 
   of neural plasticity occurring given some event. Using ordinary differential equations, neural firing equations and spike-train 
   statistics, we show how an additive short-term memory (STM) equation can be formulated
   to approach the computation of neural plasticity. We then show how the additive STM equation 
   can be used for probabilistic inference in computable neural plasticity, and the computation of 
   probabilistic neural plasticity. We will also provide a brief introduction to the theory of probabilistic
   neural plasticity and conclude with showing how it can be applied to behavioural science, 
   machine learning, artificial intelligence and psychiatry. 

\end{abstract}

\section{Introduction}

   Changes in the functionality and development of the brain has and continues to be
   an ongoing discussion and research within the neuroscience, psychology, artificial
   intelligence, cognitive science, and psychiatric community. Discoveries about the
   brain have led to major realizations about it, shifting how we act, learn, perceive and
   approach situations and circumstances. The realization of neural plasticity is one of
   those realizations about the brain that created a drastic shift in our understanding 
   of the brain as it showed us that the development of the brain does not stop at
   adulthood; rather it continues to change throughout a person’s entire life. In other 
   words, changes to the brain can take place from infancy to adulthood. 

   While the realization took place many years later, the use of it took place in the
   20th century by neuroscientist and pathologist Santiago Ramón y Cajal. He described
   neuronal plasticity as the nonpathological changes in the structure of the brain of an
   individual (Fluchs\ \& Flügge, 2014). Arguments and evidence overtime proved the
   existence of neural plasticity, leading to the current debates and discussions that we
   have about it today. Understanding the probability of neural plasticity occuring given 
   some event is a topic of interest, but has not been adequately covered or discussed. 
   To contribute to this discussion, we will propose and briefly introduce the theory of 
   probabilistic neural plasticity, then derive a short-term memory (STM) equation that can 
   be used for the computation of probabilitistic neural plasticity, elaborate how the STM 
   equation can be used for probabilistic inference in computable neural plasticity. In addition,
   we will also review relevant work on neural plasticity. We will conclude a discussion on the 
   application of probabilistic neural plasticity in areas such as behavioral science, machine learning, 
   and psychiatry.

\section{Preliminaries}

   In understanding changes within the brain, we define several terms within neuroanatomy. 
   Neurons are the cells within the nervous system that communicate with each other to transmit
   information to other nerve cells, muscles, or gland cells. Neurons generally contain an axon, 
   a cell body, and dendrites. The cell body contains the nucleus and cytoplasm. The axon is part 
   of the cell body, extending from the cell body to the nerve terminals. They often create smaller 
   branches prior to ending at the nerve terminals. Dendrites extend from the cell body and receive 
   messages from other neurons within the brain. They are composed synapses formed by the ends 
   of the axon from other neurons. 

\subsection{Synapses}

   Synapses are the spaces found within the brain cells (Bate\ \& Bassiri, 2016). Historically
   synapses were generally understood to carry the role of transferring information between one 
   neuron to another through connecting them together. However, it was later discovered that the 
   efficacy of synaptic transmission is not constant; that it varied depending on the frequency 
   of the stimulation, and the modulation of synaptic frequency provoked modifications of neural 
   connections in volume, size and shape (Bate\ \&  Bassiri, 2016). In other words, those connections 
   that are frequently stimulated will increase in volume and size, resulting in a change in shape.
   Consequently, inactive connections decrease in size and volume, resulting in synaptic long-term 
   depression. The creation of synapses is known as synaptic formation and elimination of synapses is 
   known as synaptic pruning. While synaptic formation and pruning occur extensively during the early 
   stages of brain development (i.e. embryonic brain stage), synaptic formation and pruning takes 
   place throughout until the end stages of brain development (i.e. adult brain). The adult brain 
   continues to create synapses and eliminate unwanted synapses over the course of our life in 
   response to our actions do and experiences, playing important roles in learning, 
   memory, and other aspects of the functioning of our brain. Such evolution of nervous systems 
   enables us to adapt to environments and determine the optimal action in any given situation based 
   on what was learned from our past experiences (Constandi, 2016). As a result, synapses have their 
   efficacy reinforced or weakened as a function of experience. The occurrence of modulation of synaptic 
   frequency provokes modifications of neural connections in volume, size and shape is known as neural
   plasticity, synaptic plasticity, or neuroplasticity.

\subsection{Stimuli}

   Stimuli are defined as events or occurences in the environment of an organism that influences its 
   behavior. In general, neurons responding to sensory stimuli face the challenge of encoding parameters 
   that can vary over a vast dynamic range (Dayan\ \& Abbott, 2001). To address such wide-ranging stimuli, 
   sensory neurons often respond most strongly and rapid changes in stimulus properties and are relatively 
   insensitive to steady-state levels. Mathematically, the functions of stimulus intensity are typically with 
   logarithmic or weak power-law dependence (Dayan\ \& Abbott, 2001).

\subsection{Plasticity}

   The term plasticity can be defined as a structure that is weak enough to submit to influence, but strong enough 
   to not submit to all influence at once (Constandi, 2016). From this definition, if we consider the brain as a 
   structure, the term neural plasticity can be understood as influences causing a modification of neural connections 
   with changes occurring over a specific span of time. The type of plasticity that we are
   concerned with is functional plasticity. Functional plasticity involves changes in some physiological aspect of
   nerve cell functions such as frequency of nervous impulse or the probability of release of chemical signals - 
   both of which act to make synaptic connections either stronger or weaker, resulting in changes to the degree 
   of synchronicity among populations of cells. All types of plasticity that the brain can undergo enables the brain 
   to learn from experience in order to form memories and acquire new skills, and also allows for adaptation and 
   recovery from brain trauma, or at least to compensate for and work around any damage that has occurred. 
   Consequently, the relationship between brain and behavior is not one-sided – experiences and behaviors induce 
   plastic changes in the brain, which these changes can influence our future behavior and experiences (Constandi, 
   2016). 

\section{Related work}
   Work intersecting probability theory and neural plasticity have investigated neural plasticity through using a Bayesian
   approach and probabilistic inference, resulting in the development of models. However, there are no formal theories
   that fully combine the two. The only theory that would be considered most closely related to these would be the theory of
   \textit{probabilistic neural network} (PNN) (Specht, 1989). This theory is derived from Bayesian computing
   and Kernal density method (Sebastian et al., 2019), and only focuses on artificial neural networks (ANN) (Mohebali et al., 2020).
   Aside from PNN, there is one work that is closely related that investigates neurons and behavior with respect to Bayesian inference.
   Darlington et al. have described a complete neural implementation of Bayesian-like behavior that includes an adaptation 
   of a prior. Their work focused on addressing a specific aspect of neurons - whether it is possible to decode the behavioral 
   effects of context and stimulus contrast from the neural population (Darlington et al., 2018). 

   There are some studies within the area of machine learning that propose models involving neural plasticity and probability theory.
   In a study conducted by Tully et al., they investigated the a Hebbian learning rule 
   for spiking neurons inspired by Bayesian statistics. They found that that neurons can represent information in the form 
   of probability distributions and that probabilistic inference could be a functional by-product of coupled synaptic and 
   nonsynaptic mechanisms operating over multiple timescales (Tully et al., 2014). Similarly, in a study by Kappel et al., they
   proposed that inherent stochasticity enables synaptic plasticity to carry out probabilistic inference, and presented a model 
   for synaptic plasticity that allows spiking neural networks to compensate continuously for unforeseen disturbances (Kappal 
   et al., 2015). These models help with establishing the connection between neurons and probability distributions, and synapses and
   probabilistic inference. 

\section{Probabilistic neural plasticity}

   Within plasticity, the way that probability is involved comes from the influence. More specifically, when we consider
   a structure that is weak enough to submit to influence, but strong enough to not submit to all influence at once, there
   are two probabilities that can be discussed. The first probability that can be discussed is the probability that 
   the influence can cause a structure to be influenced. The second probability that can be discussed is the probability
   that a structure is strong enough to not submit to all influence at once. This tells us that probability plays a major role
   in neural plasticity. Through understanding the role that it plays in neural plasticity; we can understand the chance 
   of neural plasticity occurring given some event. To comprehend the probability that of neural plasticity occurring given 
   some event, we define and refer to the theory of probabilistic neural plasticity.  
   
   \textbf{Definition:} Probabilistic neural plasticity is the probability that specific effects are due to causes that are 
   defined by variations of neural connections formed by the brain from adapting to its environment. 

   In this context, effects are either a behavioral condition or a mind condition. The environment consists of not only
   the physical environment, but also events, situations and circumstances. For instance, a traumatic event can be 
   considered as an environment which can cause the brain to undergo neurological changes needed for it to adapt to 
   the traumatic event. Alternatively, an environment can also be a data set as a data set used for training and testing 
   the machine learning algorithm can influence the behavior and performance of the algorithm. What can be noticed about
   probabilistic neural plasticity is that its application can be found in both human learning and in machine learning. We will
   further discuss about its application after discussing the computational components associated with probabilistic neural 
   plasticity and the computation of probabilistic neural plasticity. 

   To perform the computation of probabilistic neural plasticity, we must formulate computational and mathematical 
   models using theories and concepts from areas such as theoretical neuroscience, mathematics, and computer science.
   Furthermore, probabilistic inferences can be made through the computational aspect of neural plasticity. Moving forward,
   we will discuss and elaborate on the computation of neural mechanisms, covering neural state equations, neural firing, 
   spike-train statistics, and the neural plasticity equation. We will then discuss about probabilistic inference in computable 
   neural plasticity followed by the computation of probabilistic neural plasticity.

\section{Computable neural mechanisms and equations}

   In approaching the computation of neural mechanisms and the computation of neural plasticity, we will describe 
   a computational model of neural systems. More specifically, we will describe the computational model of neuron states, 
   neural firing, and spike-train statistics. Between physiological, psychological and neurological approaches, while the 
   computational (and mathematical) modeling can be done using any one of those approaches, we will be approaching 
   neural mechanisms and neural plasticity solely through a neurological approach. In using theoretical neuroscience and
   additive short-term memory (STM) equation, we will formulate the neural plasticity equation and a computational model
   for neural plasticity. 

\subsection{Neural state equations}

   The computational and mathematical model of neuron states can be written as neuron state equations. Assume a 
   neuron interacting with other neurons and external stimuli. Let \textit{v\textsubscript{i}} be defined as the i\textsuperscript{th}   
   neuron. Let \textit{x\textsubscript{i}} and \textit{w\textsubscript{i}} be defined as the variable that describes the neuron’s
   state. Let $\tau$ be defined as time interval. Then the one state variable of \textit{v\textsubscript{i}} is \textit{x\textsubscript{i}} 
   where \textit{x\textsubscript{i}} is denoted as the activation level of the  i\textsuperscript{th} neuron. The second state 
   variable, \textit{w\textsubscript{i}}, is associated with \textit{v\textsubscript{i}}’s interaction with \textit{v\textsubscript{j}}’s 
   (another neuron). Using additive STM equation we formulate the set of ordinary differential equations for \textit{x\textsubscript{i}} 
   and \textit{w\textsubscript{i}}. Assume a change caused by internal and external processes in neuron potential from equilibrium, and that inputs from other
   neurons and stimuli are additive (agrees with many experiments). Then for all \textit{i = 0, 1, 2, … , n}, the additive STM equation is
\begin{equation} \label{eq1}
 \mathrm{\Delta}x_i = \left(\mathrm{\Delta}x_i\right)_{internal} + \left(\mathrm{\Delta}x_i\right)_{excitatory} + \left(\mathrm{\Delta}x_i\right)_{inhibitory} + \left(\mathrm{\Delta}x_i\right)_{stimuli} \\
\end{equation}
   The additive STM equation is the $\Delta$\textit{x\textsubscript{i}} neuron state equation that 
   describes the activation level of \textit{x\textsubscript{i}} with respect to internal, excitatory, inhibitory and stimuli. We will 
   describe neural firing through the relationship between stimulus and firing-rate, and we will address the internal ordinary differential
   equation, excitatory ordinary differential equation, inhibitory ordinary differential equation, and stimuli ordinary differential equation.
   We will then formulate the neural plasticity equation needed for the computation of probabilistic neural plasticity. 

   To obtain the internal ordinary differential equation in the additive STM equation, we assume that the neuron processes are stable. In doing
   so, this provides us the internal ordinary differential equation 
\begin{equation} \label{eq2}
 \left(\mathrm{\Delta}x_i\right)_{internal} = \left(\frac{dx_i(t)}{dt}\right)_{internal} = -A_i(x_i)x_i
\end{equation}
   with \textit{A\textsubscript{i}(x\textsubscript{i})} $>$ 0. \textit{A} is denoted as the learning rates for the potentiation. 

   For the inhibitatory ordinary differential equation in the additive STM equation, it can be obtained through the change in the neuron's state.
  With \textit{w\textsubscript{i}} being defined as the variable that describes the neuron’s state, the change in the neuron's state can
   be described using $\Delta$\textit{w\textsubscript{i}}. To obtain the equation for $\Delta$\textit{w\textsubscript{i}}, we can make 
   use of existing mathematical models. A mathematical model that we can make use of is the mathematical model Bi-Phasic Spike Timing  
   Dependent Plasticity (Bi-Phasic STDP). Bi-Phasic STDP consists of two phases. The first phase is a depressive phase in which pre-
   synaptic spike follows post-synaptic spike. The second phase is a potentiating phase in which a post-synaptic spike follows pe-synaptic 
   spike (Chrol-Cannon, and Jin, 2014). The Bi-Phasic STDP model is
\begin{equation} \label{eq3}
    \mathrm{\Delta}w_i = \left\{
                \begin{array}{ll}
                 A_+exp\left\{\frac{\mathrm{\Delta}t_i}{\tau_+}\right\}\ \ \ \ \ \ \ \ \ \ \ \ \ \ \ \ \ \ \ \ \ \ \ \ \ \ \ if\ \mathrm{\Delta}t_{i\ } < 0\\
                 {-A}_-exp\left\{\frac{-\mathrm{\Delta}t_i}{\tau_-}\right\}\ \ \ \ \ \ \ \ \ \ \ \ \ \ \ \ \ \ \ \ \ if\ \mathrm{\Delta}t_{i\ } > 0
                \end{array}
              \right.
\end{equation}
   In the model, \textit{A\textsubscript{+}} is denoted as the learning rates for the potentiation and \textit{A\textsubscript{-}} is 
   denoted as the learning rates for the depression, $\Delta$\textit{t\textsubscript{i}} is denoted as the delay of the post-synaptic spike 
   occurring after the transmission of the pre-synaptic spike, and \textit{$\tau$\textsubscript{+}} and \textit{$\tau$\textsubscript{-}} 
   controls the rates of the exponential decrease in plasticity across the learning window. Using the Bi-Phasic STDP model, we obtain the 
    inhibitory ordinary differential equation 
\begin{equation} \label{eq3}
 \left(\mathrm{\Delta}x_i\right)_{inhibitory}\ =\left(\frac{dx_i(t)}{dt}\right)_{inhibitory} = \mathrm{\Delta}w_i\
\end{equation}
which
\begin{equation} \label{eq4}  
    \mathrm{\Delta}w_i = \left\{
                \begin{array}{ll}
                 A_+exp\left\{\frac{\mathrm{\Delta}t_i}{\tau_+}\right\}\ \ \ \ \ \ \ \ \ \ \ \ \ \ \ \ \ \ \ \ \ \ \ \ \ \ \ if\ \mathrm{\Delta}t_{i\ }<0\\
                 {-A}_-exp\left\{\frac{-\mathrm{\Delta}t_i}{\tau_-}\right\}\ \ \ \ \ \ \ \ \ \ \ \ \ \ \ \ \ \ \ \ \ if\ \mathrm{\Delta}t_{i\ }>0
                \end{array}
              \right.
\end{equation}

\subsection{Stimulus}

  To recall, when neurons encounter wide-ranging stimuli,  sensory neurons often respond most strongly and rapid changes
   in stimulus properties and are relatively insensitive to steady-state levels. Steady-state responses are highly compressed functions of
   stimulus intensity. Let \textit{s} be denoted as a stimulus. Applying Weber's law tells us that when we differentiate between the intensity 
   of two stimuli,  $\Delta$\textit{s} is proportional to the magnitude of \textit{s}, resulting in $\Delta$\textit{s/s} being constant. In terms 
   of adaptation, sensory systems make many adaptations, using a variety of mechanisms to adjust to the average level of stimulus intensity
   (Dayan, \ \& Abbott, 2001). When a stimulus generates such adaptation, describing responses to fluctuations about a mean stimulus level 
   is an ideal way of comprehending the relationship between the stimulus and the response (Dayan\ \& Abbott, 2001). In this case, \textit{s(t)}
    be defined such that its time average over the duration of a trial is 0. Then, we find that
\begin{equation} \label{eq5}  
  \int_{0}^{T}\frac{\left[\frac{ds(t)}{dt}\right]}{T} = 0
\end{equation}
   We can now use two approaches that we can use to progress in analyzing stimulus. The first approach would be to use several
   different stimuli and averaging over them. The second approach is putting all the stimuli that we wish to consider into a single time
   dependent stimulus sequence and average over time.  We will use the second approach, replacing stimulus averages with time
   averages. To make the stimulus periodic, for any time $\tau$, we define the stimulus outside the time limits of the trial by the relation
\begin{equation} \label{eq6}  
   s(T\ +\ \tau)\ = \ s(\tau)
\end{equation}

\subsection{Neural firing}

   For neural firing, we define spike-triggered average stimulus as \textit{C($\tau$)}, where \textit{C($\tau$)} is denoted as the 
   average value of the stimulus a time interval $\tau$ before a spike is fired. Mathematically, we define the spike-triggered average
   stimulus as 
\begin{equation} \label{eq7}  
 C\left(\tau\right) = [\frac{1}{n}\sum r(t\textsubscript{i} - \tau)] \approx [\frac{1}{<n>}\sum s(t\textsubscript{i} - \tau)]
\end{equation}
   The approximate equality of the mathematical equation comes from the fact that if \textit{n} is large, then the total number of spikes
   on each trial is well approximated by the average number of spikes per trial (Dayan, and Abbott, 2001). Expressing the spike-triggered 
   average stimulus as an integral would be
\begin{equation} \label{eq8}
 C\left(\tau\right) = \frac{1}{<n>}\int_{0}^{T}\frac{dg\left[\sum_{i = 1}^{n}{t - t_i}\right]}{dt}s\left(t + \tau\right)
\end{equation}

   In formulating the stimuli ordinary differential equation in the additive STM equation, we use the approach that consists of placing all of the
   stimuli that we want to consider into a single time-depeendent stimulus sequence and average over them - replacing stimulus averages
   with time averages. Then, for any \textit{h}, the integrals involving the stimulus being time-translationally invariant is
\begin{equation} \label{eq9}  
  \int_{0}^{T}\frac{dh\left[s(t + \tau)\right]}{dt} = \int_{\tau}^{T + \tau}\frac{dh\left[s(t)\right]}{dt} = \int_{0}^{T}\frac{dh\left[s(t)\right]}{dt}
\end{equation}
   Then the stimuli ordinary differential equation is 
\begin{equation} \label{eq10}  
 \left(\mathrm{\Delta}x_i\right)_{stimuli} = \left(\frac{dx_i(t)}{dt}\right)_{stimuli} = \int_{0}^{T}\frac{dh\left[s(t + \tau)\right]}{dt}
\end{equation}

\subsection{Spike-train statistics}

   In approaching the relationship between a stimulus and a response on a stochastic level, we use probability theory to address 
   and approach spike tunes and occurences, and neuronal firing. For spike tunes, as they are continuous variables, the probability
   for a spike to occur at a specific time is 0. In turn, to obtain a nonzero spike value, we must evaluate the probability that a spike 
   occurs within a specific time interval (Dayan\ \& Abbott, 2001). Let the probability that a spike occurs in the time interval between times \textit{t} and 
   \textit{t} + $\Delta$\textit{t}. Given such time interbal, the probability \textit{P(t\textsubscript{1},t\textsubscript{2}, ..., t\textsubscript{n})} 
   that a sequence of \textit{n} spikes occurs with spike \textit{i} falling between time interval \textit{t}\textsubscript{i} and 
   \textit{t}\textsubscript{i} + $\Delta$\textit{t} for \textit{i = 1, 2, ..., n} is given in terms of density by the relation of 
\begin{equation} \label{eq11}
  P(t_1,t_2,\ ...,\ t_n) = \ p(t_1,t_2,\ ...,\ t_n)\ \left(\Delta t\right)
\end{equation}
   To obtain an approximation of stochastic neuronal firing, Poisson process can be used as Poisson process entails that events are
   statistically independent if they are not dependent at all on preceding events (Dayan\ \& Abbott, 2001). In using Poisson process, 
   we can approach firing process through two separate cases. The first case is a homogeneous process that involves the firing rate being 
   constant over time (Dayan, \ \& Abbott, 2001). In this case, since the firing rate is constant, the Poisson process generates every sequence 
   of n spikes over a fixed time interval with equal probability. Let \textit{P\textsubscript{T}(n)} be the probability that an arbitrary sequence of 
   exactly n spikes occurring within a trial of duration \textit{T}. Let the firing rate for a homogeneous Poisson process be denoted as\textit{q(t) = q)} 
   (as it is independent of time). Assume that the spike times are ordered. Then \textit{0 $\leq$ t\textsubscript{1} $\leq$ t\textsubscript{1} $\leq$ ... $\leq$ t\textsubscript{n} $\leq$ T} 
   and the probability for \textit{n} spike times is
\begin{equation} \label{eq12}
 P(t_1,t_2,\ ...,\ t_n) = n!P_T\left(n\right)\left(\frac{\Delta t}{T}\right)^n
\end{equation}
   The second case is a inhomogeneous poisson process that involves the firing rate being time-dependent (Dayan, \ \& Abbott, 2001).  In this 
   case, since the firing rate depends on time, different sequences of \textit{n} spikes occur with different probabilities. In addition, the probability  
   \textit{p(t\textsubscript{1},t\textsubscript{2}\ ...,\ t\textsubscript{n})} depends on the spike times. Using inhomogeneous Poisson 
   process, we find that the spikes are still generared independently, and their times enter \textit{p(t\textsubscript{1},t\textsubscript{2}\ ...,\ t\textsubscript{n}} 
   only through the time-dependent firing rate \textit{r(t)}. Assume that the spike times are ordered. Then \textit{0 $\leq$ t\textsubscript{1} $\leq$ t\textsubscript{1} $\leq$ ... $\leq$ t\textsubscript{n} $\leq$ T}
   and the probability for \textit{n} spike times is 
\begin{equation} \label{eq13}
 p(t_1,t_2,\ ...,\ t_n) = \prod_{i = 1}^{n}r\left(t_i\right)exp\left\{\int_{0}^{T}\frac{dr\left(t\right)}{dt}\right\}
\end{equation}
   For the excitatory ordinary differential equation in the additive STM equation, we assume additive synaptic excitation proportions to the spike-train frequency. 
   We find that for the excitatory ordinary differential equation, the equation depends on the firing rate and whether it is constant over
   time or time-dependent. This gives us the ordinary differential equation 
\begin{equation} \label{eq14}
 \left(\mathrm{\Delta}x_i\right)_{excitatory} = \left(\frac{dx_i(t)}{dt}\right)_{excitatory} = \rho_i
\end{equation}
 where
\begin{equation} \label{eq15}
    \rho_i = \left\{
                \begin{array}{ll}
                  n!P_T\left(n\right)\left(\frac{\Delta t}{T}\right)^n\ \ \ \ \ \ \ \ \ \ \ \ \ \ \ \ \ \ \ \ \ \ \ if\ the\ firing\ rate\ is\ constant\ over\ time\\
                  \prod_{ i = 1}^{n}r\left(t_i\right)exp\left\{\int_{0}^{T}\frac{dr\left(t\right)}{dt}\right\}\ \ \ \ if\ the\ firing\ rate\ is\ time-dependent
                \end{array}
              \right.
\end{equation}

\subsection{Neural plasticity equation}

   Combining all of the ordinary differential equations together, for all \textit{i = 0, 1, …. n}, the neuron state equation is
\begin{equation} \label{eq16}
\begin{aligned}
 \mathrm{\Delta}x_i &= \left(\mathrm{\Delta}x_i\right)_{internal} + \left(\mathrm{\Delta}x_i\right)_{excitatory} + \left(\mathrm{\Delta}x_i\right)_{inhibitory} + \left(\mathrm{\Delta}x_i\right)_{stimuli} \\
                                 &= -A_i(x_i)x_i\ + \rho_i + \mathrm{\Delta}w_i\ + \int_{0}^{T}\frac{dh\left[s(t+\tau\right)]}{dt}
\end{aligned}
\end{equation}
   which \textit{$\rho$\textsubscript{i}} and \textit{w\textsubscript{i}} are defined with respect to their ordinary differential equations.
   To compute neural plasticity, we consider all of the activities of the neurons from 1 to n. The activity level that most frequently 
   occurred among all the neurons is the neuron state with has the highest degree of neural plasticity. The neuron state that has the 
   highest degree of neural plasticity therefore has the most influence on the brain, directly contributing to the (plastic) changes 
   occurring inside the brain. The final computational model that can be used for neural plasticity, which we will name the Degree of 
   Neural Plasticity Model (DNP model) would then become the following: let \textit{x\textsubscript{i}} and \textit{w\textsubscript{i}} be 
   defined as the variables that describe the neuron’s state. Then for all \textit{i = 0, 1, …. n} 
\begin{equation} \label{eq17}
 DNP\ = Mode([\mathrm{\Delta}x_i]) = Mode([\mathrm{\Delta}x\textsubscript{1},\mathrm{\Delta}x\textsubscript{2},...,\mathrm{\Delta}x\textsubscript{n}])
\end{equation}
   In the model, we denote $\Delta$\textit{w\textsubscript{i}} as the change in synaptic strength; \textit{w\textsubscript{i}} as the 
   synaptic strength; \textit{A\textsubscript{+}} as the learning rate for the potentiation; \textit{A\textsubscript{-}} as the learning 
   rate for the depression; \textit{$\tau$\textsubscript{+}} and \textit{$\tau$\textsubscript{-}} as what controls the rates of 
   exponential decrease across the learning window; T as trial; and \textit{x\textsubscript{i}} as the activation level of the
   i\textsuperscript{th} neuron. This establishes the computation of neural plasticity.

\section{Probabilistic inference in computable neural plasticity}

   With establishing the computation of neural plasticity, we can discuss how probabilistic inferences can be made through the 
   computation of neural plasticity. Probabilistic inference is used to approach probabilistic queries of form \textit{P$(X|Y)$}, where 
   \textit{Y} and \textit{Z}, are disjoint subsets of \textit{X} given a graphic model for \textit{X}. We will go into the 
   construction of the graphical model. Rather, we will directly define a probabilistic query and address the probabilistic queries using 
   probabilistic inference. However, note that a basic graphical model can be constructed using the concept of neuron state equations
   and the Bi-Phase STDP model. In understanding neural plasticity, our query would be to know what the posterior distribution of degrees 
   of neural plasticity would look like. Let \textit{Y} be the query variable. Let \textit{T} be denoted as a time interval \textit{[t\textsubscript{i}, t]} 
   with \textit{i = 1, 2, ..., n}. Let \textit{DNP} be denoted as the degrees of neural plasticity. The probabilistic query can address using is 
\begin{equation} \label{eq17}
\begin{aligned}
  P\left(Y\middle| T,\ DNP\right) &= P(Y|DNP_1,\ DNP_2,\ ...,DNP_n) \\
                                 &= P[Y|DNP(\ T_1\ =\ t_1),\ DNP(\ T_2\ =\ t_2),\ ...,\ DNP(\ T_n\ =\ t_n)]
\end{aligned}
\end{equation}
   This would enable us to obtain the posterior probabilities of degree of neural plasticity. This in turn, can then be used to estimate the 
   posterior distribution with respect to degrees of neural plasticity.

\section{Computation of probabilistic neural plasticity}

   To recall, the problem that is being answered is what is the probability that event \textit{X} will cause neural plastic changes to the 
   brain. The answer to that question can be obtained through using probability neural plasticity - the probability that specific effects are 
   due to causes that are defined by variations of neural connections formed by the brain from adapting to its environment. With the 
   DNP model, we can now investigate and analyze how exactly the computation of probabilistic neural plasticity can be achieved.  The 
   first thing we will do is formulate the problem in mathematical terms.
   Let E be denoted as a specific event Let T be denoted as a trial with \textit{[t\textsubscript{i}, t]} with \textit{i = 1, 2, ..., n}. Let 
   \textit{DNP} be denoted as the degrees of neural plasticity. The problem in mathematical terms would be \textit{P$(DNP|E,T)$}. The 
   variable \textit{T} was added as the time interval must be defined in order to compute the change in synaptic strength. The 
   probabilistic neural plasticity model that would be 
\begin{equation} \label{eq18}
 P\left(DNP\middle|E,\ T\right) = \frac{P(E|DNP,\ T)P(T|DNP)P(DNP)}{P(DNP|E,\ T)P(E|T)P(T)}
\end{equation}
   Since\ T must always be known, \textit{P(T) = 1}. For \textit{P(DNP)}, since it relies on trial \textit{T}, we cannot compute for 
   \textit{P(DNP)} directly.  To address this, we can either assume that \textit{P(DNP) = 1} or \textit{P(DNP) = 0}, which for this case 
   we will always assume that \textit{P(DNP) = 1}.  The only time when \textit{P(DNP) = 0} is when $\Delta$\textit{t\textsubscript{i} =} 0. 
   This simplifies probabilistic neural plasticity model \textit{P$(DNP|E,T)$} to 
\begin{equation} \label{eq19}
 P\left(DNP\middle| E,\ T\right) = \frac{P(E|DNP,\ T)P(T|DNP)}{P(DNP|E,\ T)P(E|T)}
\end{equation}
   Set DNP as the function with respect to trial T. Then
\begin{equation} \label{eq20}
  DNP\ = Mode([\mathrm{\Delta}x_i]) = Mode([\mathrm{\Delta}x\textsubscript{1},\mathrm{\Delta}x\textsubscript{2},...,\mathrm{\Delta}x\textsubscript{n}]) 
\end{equation}
   which this can be solved using the STM equation. From here, we can compute the probability of whether stimuli \textit{S} will most likely cause neural 
   plastic changes to the brain. If a high probability is obtained from the computation of \textit{P$(DNP| S,T)$}, then that implies that 
   stimuli \textit{S} will most likely cause neural plastic changes to the brain. If a low probability is obtained from the computation of 
   \textit{P$(DNP| S,T)$}, then that implies that stimuli \textit{S} will not most likely cause neural plastic changes to the brain.

\section{Application of probabilistic neural plasticity}

   Our theoretical results stems from behavioral science and machine learning. We wish to study the behavior of humans and artificial
   intelligence (AI), understanding what is the probability of neural plasticity occuring given some event. For humans, the event would be 
   a stimulus, while for artificial intelligence, the event would be data. In both cases, we wish to understand how likely it is for a human or 
   machine learning algorithm to change or develop a new behavior from experiencing some event. Another way of seeing it is that we 
   want to understand how likely it is for a human or machine learning algorithm to change its thought and reasoning process after 
   experiencing some event. The cases presented are not actual cases that have taken place. Rather, they are used to illustrate how 
   probabilisitic neural plasticity can be applied to behavioral science, machine learning, and psychiatry. In addition, it is important to note
   that while our dicussion focuses on results that involve behavioral science, machine learning and psychiatry, discussion on the results 
   (and even the application of probabilistic neural plasticity) can futher be elaborated, extending and involving other disciplines such as 
   neuroscience and cognitive science. 

\subsection{Behavioral science and psychiatry}

   Within the context of behavioral science, we consider a case in psychiatry. A patient diagnosed with borderline personality disorder is 
   presented. It is provided that the patient expressed chronic feelings of emptiness, and has a fear of abandonment. In addition, the 
   patient display feelings of anger, but does not provide adequate details about their relationship with their parents, making it difficult to 
   determine whether the cause of the chronic feelings of emptiness and fear of abandonment is due to the patient being neglected,  
   withheld, uncared, or abandoned by the caregivers (American Psychiatric Association, 2013). To determine whether neglect from the 
   parents caused neural plastic changes to the brain of the patient, we apply probabilistic neural plasticity. Let \textit{DNP} be defined 
   as the degree of neural plasticity with respect to the behavior and feelings presented by the patient (chronic feelings of chronic 
   feelings of emptiness, fear of abandonment, and anger). Let \textit{E} be defined as the parents action towards the patient, and let 
   \textit{T} be the time interval. Then the equation (20) becomes
\begin{equation} \label{eq21}
\begin{aligned}
  P\left(DNP\middle| E,\ T\right) &= P\left(DNP\middle| Action_{parents},\ T\right)  \\
   				        &= \frac{P(Action_{parents}|DNP,\ T)P(T|DNP)P(DNP)}{P(DNP|Action_{parents},\ T)P(Action_{parents}|T)P(T)}
\end{aligned}
\end{equation}
   We set the DNP as a function with respect to trial T. Then
\begin{equation} \label{eq22}
  DNP\ =Mode([\mathrm{\Delta}x_i])=Mode([\mathrm{\Delta}x\textsubscript{1},\mathrm{\Delta}x\textsubscript{2},...,\mathrm{\Delta}x\textsubscript{n}]) 
\end{equation}
   Let \textit{n} be the current age of the patient. We expand equation (16) with \textit{$\rho$\textsubscript{i}} being defined with the 
   firing rate being time-dependent, and $\Delta$\textit{t\textsubscript{i}} $\geq$ 0. Then the equation for \textit{i = 1, 2, ..., n}, $\Delta$\textit{x\textsubscript{i}} is
\begin{equation} \label{eq23}
\begin{aligned}
 \mathrm{\Delta}x_i &= -A_i(x_i)x_i\ + \rho_i + \mathrm{\Delta}w_i\ + \int_{0}^{T}\frac{dh\left[s(t+\tau\right)]}{dt} \\
                                 &= -A_i(x_i)x_i\ + \prod_{i=1}^{n}r\left(t_i\right)exp\left\{\int_{0}^{T}\frac{dr\left(t\right)}{dt}\right\} + {-A}_-exp\left\{\frac{-\mathrm{\Delta}t_i}{\tau_-}\right\} + \int_{0}^{T}\frac{dh\left[s(t+\tau\right)]}{dt}
\end{aligned}
\end{equation}
   If the probability \textit{P$(DNP| E,T)$} $\geq$ 0.5, then neglect from the parents caused neural plastic changes to the brain of the 
   patient. If \textit{P$(DNP| E,T)$} $<$ 0.5, then neglect from the parents did not cause neural plastic changes to the brain of 
   the patient. 

\subsection{Machine learning and artificial intelligence}

   Within the context of machine learning, we consider an ANN. An ANN designed to predict the probability
   that an individual that has committed a crime in the past will commit another crime in the future. We know that the data used to train
   the ANN was provided by the department of justice of the country. The data included sex (male, female), age at release (18 and 
   older) and race/hispanic origin (white, black/African American, hispanic/latino, others), commitment offense (violent, property, drug,
   and public order), and recidivism. We also know that based on the data, previously convicted males of age range 18 - 29 of race 
   others that commit an offense of violence are most likely to recommit the act of violence. The ANN is being used on a white male, aged 
   24, that has previously been convicted of violence to determine the probability that that invidivual will commit the same crime again in
   the next 3 years. The AI (that uses the ANN) predicts that the individual will commit the same crime in the next 3 years. To determine
   whether the commitment offense data caused neural plastic changes to the ANN to determine whether the individual will 
   recommit the crime in the next 3 years, we apply probabilistic neural plasticity. Let \textit{DNP} be defined 
   as the degree of neural plasticity with respect to the individual's criminal record. Let \textit{E} be defined as the criminal offense, and 
   let \textit{T} be the time interval. Then the equation (20) becomes
\begin{equation} \label{eq24}
\begin{aligned}
  P\left(DNP\middle| E,\ T\right) &= P\left(DNP\middle| Offense_{violence},\ T\right)  \\
   				        &= \frac{P(Offense_{violence}|DNP,\ T)P(T|DNP)P(DNP)}{P(DNP|Offense_{violence},\ T)P(Offense_{violence}|T)P(T)}
\end{aligned}
\end{equation}
   We set the DNP as a function with respect to trial T. Then
\begin{equation} \label{eq25}
  DNP\ =Mode([\mathrm{\Delta}x_i])=Mode([\mathrm{\Delta}x\textsubscript{1},\mathrm{\Delta}x\textsubscript{2},...,\mathrm{\Delta}x\textsubscript{n}]) 
\end{equation}
   Let \textit{n} be the number of years. We expand equation (16) with \textit{$\rho$\textsubscript{i}} being defined with the 
   firing rate being constant over time, and $\Delta$\textit{t\textsubscript{i}} $<$ 0. Then the equation for \textit{i = 1, 2, ..., n}, $\Delta$\textit{x\textsubscript{i}} is
\begin{equation} \label{eq26}
\begin{aligned}
 \mathrm{\Delta}x_i &= -A_i(x_i)x_i\ + \rho_i+\mathrm{\Delta}w_i\ + \int_{0}^{T}\frac{dh\left[s(t + \tau\right)]}{dt} \\
                                 &= -A_i(x_i)x_i\ + n!P_T\left(n\right)\left(\frac{\Delta t}{T}\right)^n+ A_+exp\left\{\frac{\mathrm{\Delta}t_i}{\tau_+}\right\} + \int_{0}^{T}\frac{dh\left[s(t+\tau\right)]}{dt}
\end{aligned}
\end{equation}
   If the probability \textit{P$(DNP| E,T)$} $\geq$ 0.5, then the commitment offense in the data caused neural plastic changes to the ANN, 
   resulting in the decision made by the AI on the previously convicted individual. If \textit{P$(DNP| E,T)$} $<$ 0.5, then then the 
   commitment offense data did not cause neural plastic changes to the ANN, resulting in the decision made by the AI on the previously 
   convicted individual.

\section{Discussion}

   In attempt to study the behavior of humans and AI, understanding what is the probability of neural plasticity occuring given some event. 
   For humans, the event would be a stimulus, while for artificial intelligence, the event would be data. In both cases, we wish to understand how likely it is for a 
   human or machine learning algorithm to change or develop a new behavior from experiencing some event. Accordingly, to address this, we proposed 
   probabilistic neural plasticity, performed the computation of it, then applied it to behavioral science, psychiatry, machine learning and AI. In proposing and 
   developing the theory of probabilistic neural plasticity, this contributes to the field of neuroscience and artificial intelligence, and brings us a step closer to 
   answering questions and provide explanations pertaining to learning and behavior, especially within humans. Future work on this theory can focus on collecting 
   empirical evidence to better establish it, ellaborating on the theory itself, and applying the theory in areas such as cognitive science, psychology and deep 
   learning. 

\section{Acknowledgement}
We would like to thank Linbo Wang (University of Toronto) and Mark Fortney (University of Toronto) for useful discussions and suggestions. This manuscript has been released as a pre-print at arXiv (Hossain, 2019). 

\section*{References}

\small

American Psychiatric Association (2013). Diagnostic and Statistical Manual of Mental Disorders. \textit{American Psychiatric Publishing}. 5, 664.

Bates, D., \ \& Bassiri, N. (2016). Plasticity and Pathology. \textit{Berkeley Forum in the Humanities}. 27 – 28. 

Chrol-Cannon, J., \ \& Jin, Y. (2014). Computational Modeling of Neural Plasticity for Self-Organization of Neural Networks. \textit{Biosystems} 125, 43 – 54.

Constandi, M. (2016). Neuroplasticity. \textit{The MIT Press Essential Knowledge Series}. 2, 8 – 13, 28 – 29, 45 – 47. 

Dayan, P., \ \& Abbott, L. F. (2001). Theoretical Neuroscience – Computational and Mathematical Modeling of Neural Systems. \textit{The MIT Press}. 17 – 39, 47 – 51.

Darlington, T. R., Beck, J. M., \ \& Lisberger, S. G. (2018). Neural Implementation of Bayesian Inference in a Sensorimotor Behavior. \textit{Nature Neuroscience}. DOI: 10.1038/s41593-018-0233-y 

Fuchs, E., \ \& Flügge, G. (2014). Adult Neuroplasticity: More Than 40 Years of Research. \textit{Hindawi Publishing Corporation}. 1. 

Hossain, S. Application and Computation of Probabilistic Neural Plasticity. arXiv:1907:00689. 

Kappal, D., Habenschuss, S., Legenstein R., \ \& Maass, W. (2015). Synaptic Sampling: A Bayesian Approach to Neural Network Plasticity and Rewiring. \textit{In Proceedings of the 28th International Conference on Neural Information Processing Systems} (NeurIPS 2015). 370 - 378. 

Mohebali, B., Tahmassebi, A., Meyer-Baese, A., \ \& Gandomi, A. H. (2020). \textit{Probabilistic Neural Networks. Handbook of Probabilistic Models}. 347–350. DOI: 10.1016/b978-0-12-816514-0.00014-x 

Sebastian, A., Pannone, A., Subbulakshmi Radhakrishnan, S., \ \& Das, S. (2019). Gaussian Synapses for Probabilistic Neural Networks. \textit{Nature Communications}. 10(1). DOI: 10.1038/s41467-019-12035-6 

Specht, D. F. (1990). Probabilistic neural networks. \textit{Neural Networks}. 3(1), 109–110. DOI: 10.1016/0893-6080(90)90049-q

Tully, P. J., Hennig, M. H., \ \& Lansner, A. (2014). Synaptic and Nonsynaptic Plasticity Approximating Probabilistic Inference. \textit{Frontiers in Synaptic Neuroscience}. 6(8), 1 - 16. DOI: 10.3389/fnsyn.2014.00008

\end{document}